\documentclass[twocolumn,prb,showpacs,preprintnumbers,groupedaddress]{revtex4}
\usepackage{amssymb}
\usepackage{mathptm}
\usepackage{dcolumn}
\usepackage{bm}
\usepackage{times}
\usepackage{graphicx}

\begin{document}

\title{Interband superconductivity: contrasts between BCS and Eliashberg
theory}
\author{O.V.~Dolgov$^1$}
\affiliation{$^1$Max-Planck-Institut f\"{u}r Festk\"{o}rperforschung, D-70569 Stuttgart,
Germany}
\author{I.I.~Mazin$^{2}$}
\author{D.~Parker$^{2}$}
\affiliation{$^2$Naval Research Laboratory, 4555 Overlook Ave. SW, Washington, DC 20375}
\author{A.A.~Golubov$^{3}$}
\affiliation{$^3$Faculty of Science and Technology, University of Twente, 7500 AE
Enschede, The Netherlands}
\date{\today}

\begin{abstract}
The newly discovered iron pnictide superconductors apparently present an
unusual case of interband-channel pairing superconductivity. Here we show
that, in the limit where the pairing occurs within the interband channel,
several surprising effects occur quite naturally and generally: different
density-of-states on the two bands lead to several unusual properties,
including a gap ratio which behaves inversely to the ratio of
density-of-states; the weak-coupling limit of the Eliashberg and the BCS theory, commonly
taken as equivalent, in fact predict qualitatively different dependence of
the $\Delta_{1}/\Delta_{2}$ and $\Delta/T_{c}$ ratios on coupling constants.
We show analytically that these effects follow directly from the
interband character of
superconductivity. Our results show that in the interband-only pairing model
the maximal gap ratio is $\sqrt{N_{2}/N_{1}}$ as strong-coupling effects act only to reduce this ratio.  This suggests that if the large experimentally reported
gap ratios (up to a factor 2) are correct, the pairing mechanism must include
more intraband interaction than is usually assumed.
\end{abstract}

\maketitle

Athough first proposed 50 years ago, multiband superconductivity where the
order parameter is different in different bands had not attracted much
interest until 2001 when MgB$_{2}$ was found to be a two-band suprconductor.
MgB$_{2}$ represents a partucular case where one \textquotedblleft
leading\textquotedblright\ band enjoys the strongest pairing interactions,
while the interband pairing interaction, as well as the intraband pairing in the other band,
are weak. There is growing evidence that the newly disovered superconducting
ferropnictides represent another limiting case: the pairing interaction
is predominantly interband, while the intraband pairing in both bands is
weak. This leads to a number of interesting and qualitatively new effects,
including the fact that a repulsive interband interaction is nearly as
effective in creating superconductivity as an attractive one.

In this paper we will show another surprising feature of the two-band
\textquotedblleft interband\textquotedblright\ superconductivity (meaning
superconductivity induced predominantly by interband interactions): entirely
counterintuitively, \textit{the BCS theory for such superconductors is not
the weak coupling limit of the Eliashberg theory, }and the difference is not
only quantitative but qualitative. This fact holds for either repulsive (as,
presumably, in pnictides) or attractive interactions.

Specifically, we will concentrate on the dependence of the superconducting
gaps in the two bands on the ratio of the densities of states and the
magnitude of the superconducting coupling. We will show that the gap ratio is
always smaller in the Eliashberg theory than in the BCS theory, the
deviation grows with coupling strength and with temperature, and is
largest just below $T_{c}.$

Let us start with the BCS equations\cite{BCS}. For a two band interband-only
case, with gap parameters given on the two bands as $\Delta _{1}$ and $%
\Delta _{2}$, the BCS gap equations take the form 
\begin{eqnarray}
\Delta _{1} &=&\sum_{k}\frac{V\Delta _{2}\tanh (E_{2,k}/2k_{B}T)}{2E_{2,k}} 
\nonumber \\
\Delta _{2} &=&\sum_{k}\frac{V\Delta _{1}(k)\tanh (E_{1,k}/2k_{B}T)}{%
2E_{1,k^{^{\prime }}}}  \label{BCS}
\end{eqnarray}%
where $E_{i,k}$ is the usual quasiparticle energy in band $i$ given by $%
\sqrt{(\epsilon _{i,k}-\mu )^{2}+\Delta _{i}^{2}}$, the normal state
electron energy is $\epsilon _{i,k},$ $\mu $ is the chemical potential. and $%
V$ is the interband interaction causing the superconductivity. $V$ can be
either attractive ($>0$ in this convention) or repulsive (as presumably in the pnictides), but
for the rest of the paper the sign does not matter. For simplicity we will
use $V>0$ and $\Delta >0,$ keeping in mind that for pnictides all the
results apply by substituting $\Delta $ by $|\Delta |.$ The BCS theory assumes $V$ to
be constant up to the cut-off energy $\omega _{c}$. Following the BCS
prescription, we can convert the momentum sums to energy integrals up to a
cut-off energy $\omega _{c}$ and assume Fermi-level density-of-states (DOS) $%
N_{1}$ and $N_{2}.$ Near $T_{c}$ these equations can be linearized giving

\begin{eqnarray}
\Delta _{1} &=&\Delta _{2}\lambda _{12}\log (1.136\omega _{c}/T_{c}) 
\nonumber \\
\Delta _{2} &=&\Delta _{1}\lambda _{21}\log (1.136\omega _{c}/T_{c}),
\label{BCSTc}
\end{eqnarray}%
where $\lambda _{12}=N_{2}V$, the dimensionless coupling constant, with a
similar expression for $\lambda _{21}$ $.$ These equations readily yield $%
\lambda _{eff}=\sqrt{\lambda _{12}\lambda _{21}}$ and $\alpha =\Delta
_{2}/\Delta _{1}=\sqrt{N_{1}/N_{2}}.$ This result has been obtained before%
\cite{parker,bang}. Similarly, at $T=0$ in the weak-coupling limit 
\begin{eqnarray}
\Delta _{1} &=&\Delta _{2}\lambda _{12}\sinh ^{-1}(\omega _{c}/\Delta _{2}) 
\nonumber \\
\Delta _{2} &=&\Delta _{1}\lambda _{21}\sinh ^{-1}(\omega _{c}/\Delta _{1})
\label{BCST0}
\end{eqnarray}%
Obviously, for $\lambda _{eff}\rightarrow 0$ we have $T_{c}\rightarrow 0$
and the relation $\Delta _{2}/\Delta _{1}=\sqrt{N_{1}/N_{2}}$ should hold.
The same is not true for $\lambda _{eff}>0.$

First principle calculations suggest for the pnictides $\beta
=N_{2}/N_{1}\lesssim 1.4,$ corresponding to the gap ratio $\alpha \lesssim
1.2.$ Experimental estimates for the gaps differ wildly, yielding gap ratios
ranging from 1.3 to 3.4. Since the goal of this paper is to address the
effect of the density of states difference on the gap ratio, we will use an
intermediate number\cite{malone} $\alpha =1.6$ ($\beta =2.6).$ 
\begin{figure}[h]
\includegraphics[width=8cm]{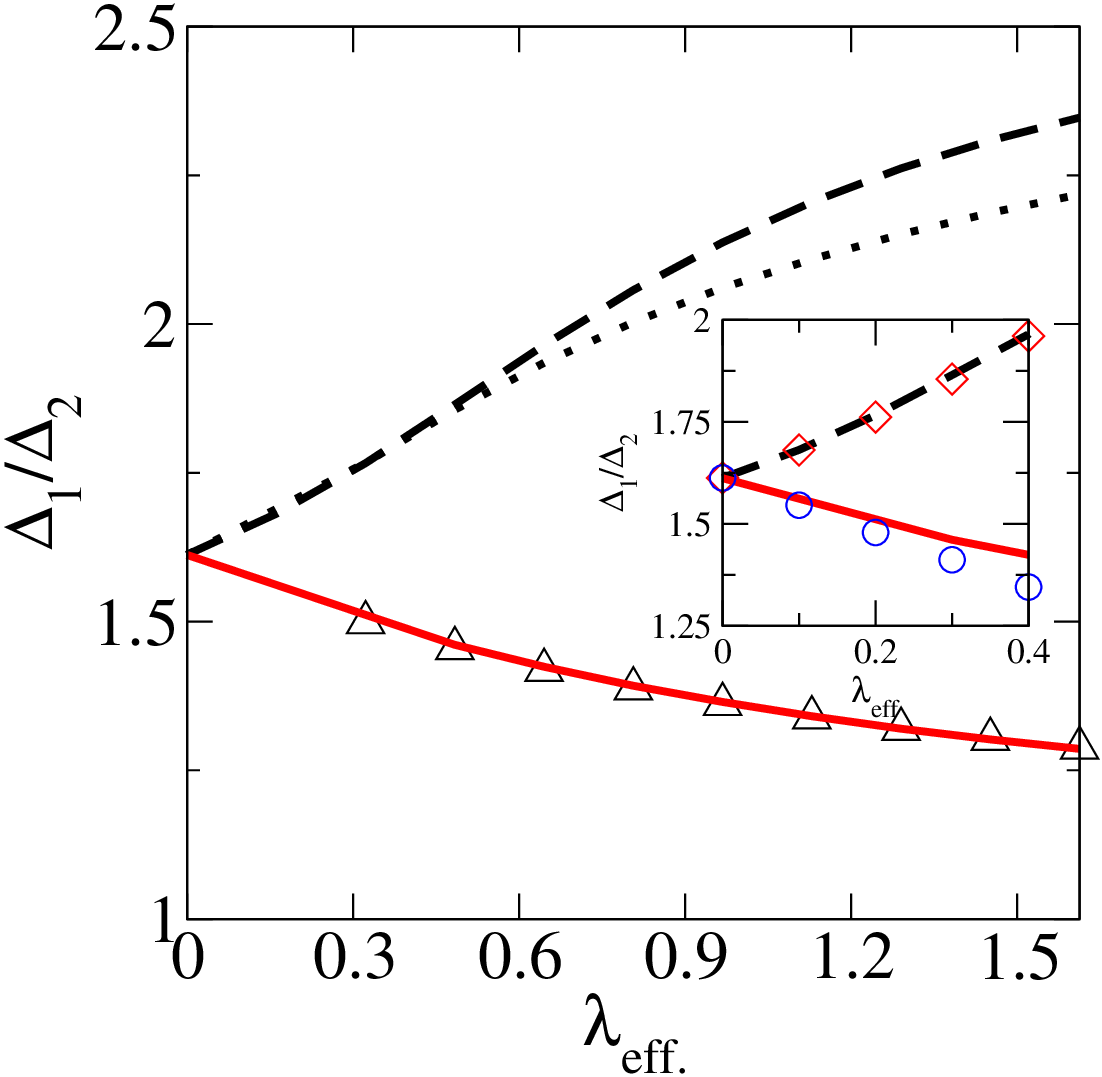}
\caption{(color online) The ratio of the gap functions in an interband
pairing case, as a function of $\protect\lambda _{eff}$, for the BCS (dashed line) and Eliashberg Einstein (line) spectrum and spin fluctuation (triangle) spectrum cases.  The dotted line represents numerical Eliashberg Einstein spectrum results in which the mass renormalization parameter has been artificially taken as 1, showing that the difference between BCS and Eliashberg is mainly a mass renormalization effect. Inset: analytic approximations to numerical results: diamonds are BCS Eq. \ref{BCS2}, circles are Eliashberg Eq. \ref{final}.}
\label{fig1}
\end{figure}

The fact that the band with the \textit{larger} DOS ends up with a \textit{%
smaller} gap is a somewhat counterintuitive result. This is a direct result of
the interband-only pairing - the pairing amplitude on one band is generated
by the DOS on the \textit{other}. Numerical solution of Eqs. \ref{BCSTc} at $%
T=0$ (Fig. \ref{fig1} gives, as expected, $\alpha =\sqrt{\beta }=1.61$ at $%
\lambda _{eff}\rightarrow 0.$ As a function of $\lambda _{eff}$ it increases
linearly, reaching $\approx 2.3$ at $\lambda _{eff}\approx 1.6$ (note that
as shown below, it will ultimately saturate at $%
\beta =2.6$ in the superstrong limit ). This increase can be
easily explained.

Let us define $\alpha =\beta ^{1/2}$, so that $x\ll 1$ at $\lambda \ll 1,$
and substitute $\sinh ^{-1}(\omega _{c}/\Delta )\rightarrow \log (2\omega
_{c}/\Delta ).$ A few lines of algebra then lead to 
\begin{equation}
x=\frac{\log {\beta }}{2(1+2/\sqrt{\lambda _{12}\lambda _{21}})}\simeq \frac{%
\lambda _{eff}\log \beta }{4}  \label{BCSa}
\end{equation}%
This result was also obtained by Bang \cite{bang}.  {The quadratic in }$\lambda 
$ term can also be worked out and reads {%
\begin{equation}
\frac{\Delta _{2}}{\Delta _{1}}=\sqrt{\beta }[1+\frac{\lambda _{eff}\log
\beta }{4}+\frac{\lambda_{eff} ^{2}(4\log
\beta +\log ^{2}\beta )}{32}] 
\label{BCS2}
\end{equation}%
As Fig. \ref{fig1} shows, this expression describes the numerical solution
at small $\lambda $ very well. Although not apparent from the plots, the $\Delta_{1}/\Delta_{2}$ ratio
will saturate at large $\lambda$, as shown by Bang and Choi \cite{bang} and can also be seen from Eq. 3,
since 
\begin{eqnarray}
\Delta _{1} &=&\Delta _{2}\lambda _{12}\sinh ^{-1}(\omega _{c}/\Delta _{2}) \rightarrow \lambda_{12}\omega_{c} \mathrm{\,\,for\,\,} \Delta_{2} \gg \omega_{c}
\end{eqnarray}%
Similarly, in this limit $\Delta_{2} = \lambda_{21} \omega_{c}$ so that
$\Delta_{1}/\Delta_{2} = \lambda_{12}/\lambda_{21}=N_{2}/N_{1}$.
All these BCS results, however are opposite to a known analytical
result \cite{GD} that in the superstrong (Eliashberg) limit $\lambda \gg 1$ the gap ratio $\alpha\rightarrow 1$ independent of $\beta $.  Let us now move to the strong-coupling
limit, given by Eliashberg \cite{eliashberg} theory.
}

In this theory, the BCS
gap function $\Delta _{0}$ is replaced by a complex, energy-dependent
quantity $\Delta _{0}(\omega )$, which must be determined along with a mass
renormalization parameter $Z(\omega )$. One commonly formulates the
equations in terms of $\phi (\omega )=Z(\omega )\Delta (\omega )$, and these
equations can be solved either on the real frequency axis or the imaginary
axis (using Matsubara frequencies). These equations are formulated in a
two-band interband pairing case on the imaginary axis as follows (some of
the notation is repeated from \cite{nicol}): 
\small
\begin{eqnarray}
\Delta _{1}(i\omega _{n})Z_{1}(i\omega _{n}) &=&\pi T\sum_{m}K_{12}(i\omega
_{m}-i\omega _{n})\frac{\Delta _{2}(i\omega _{m})}{\sqrt{\omega
_{m}^{2}+\Delta _{2}^{2}(i\omega _{m})}}  \label{El1} \\
Z_{1}(i\omega _{n}) &=&1+\frac{\pi T}{\omega _{n}}\sum_{m}K_{12}(i\omega
_{m}-i\omega _{n})\frac{\omega _{m}}{\sqrt{\omega _{m}^{2}+\Delta
_{2}^{2}(i\omega _{m})}}  \label{El2}
\end{eqnarray}%
\normalsize
Here the kernel K$_{12}$ is given by 
\[
K_{12}(i\omega _{m}-i\omega _{n})=2\int_{0}^{\infty }\frac{\Omega
B_{12}(\Omega )d\Omega }{\Omega ^{2}+(\omega _{n}-\omega _{n})^{2}} 
\]%
This ${B}_{12}$ represents the electron-boson coupling function which
supplants the pairing potential used in BCS theory, and there is an exactly analogous
equation for band 2. Here $B_{12}(\Omega )/B_{21}(\Omega )=N_{2}/N_{1}=\beta
.$

First we assume a simple Einstein-type electron-boson coupling function.
Numerical solution of the Eliashberg equations (\ref{El2}) finds that the
ratio of the gaps \textit{decreases} with $\lambda ,$ opposite to the BCS
prediction that the ratio of the gaps \textit{increases} with increasing
coupling. This can be understood analytically as well.

First of all, we observe that neglecting the mass renormalization by setting 
$Z=1$ in Eq.\ref{El1} appears to be very close to the BCS solution (in fact,
deviation from the lowest-order approximation of Eq. \ref{BCSa} is mainly
due to the increasing difference between $\sinh ^{-1}(\omega _{c}/\Delta )$
and $\log (2\omega _{c}/\Delta ))$. Let us now work out the effect of the
mass renormalization.

Assuming an Einstein spectrum with the frequency $\Omega$, at T=0 Eqs. \ref{El1},%
\ref{El2} {reduce to%
\small
\begin{eqnarray}
\Delta _{1}(\omega )Z_{1}(\omega ) &=&\frac{\lambda _{12}\Omega ^{2}}{2}\int_{-\infty
}^{\infty }\frac{d\omega ^{^{\prime }}\Delta _{2}(\omega ^{^{\prime }})}{%
(\Omega ^{2}+(\omega -\omega ^{^{\prime }})^{2})(\sqrt{\omega ^{^{\prime
}2}+\Delta _{2}^{2}(\omega )})} \nonumber
\end{eqnarray}
and
\begin{eqnarray}
Z_{1}(\omega ) &=&1+\frac{1}{2\omega }\lambda _{12}\Omega ^{2}\int_{-\infty
}^{\infty }d\omega ^{`}\frac{\omega ^{^{\prime }}}{(\Omega ^{2}+(\omega
-\omega ^{^{\prime }})^{2})(\sqrt{\omega ^{^{\prime }2}+\Delta
_{2}^{2}(\omega )})} \nonumber
\end{eqnarray}%
\normalsize
with a similar equation for $\Delta_{2}$ and Z$_{2}$. 
In the popular \textquotedblleft square-well\textquotedblright\ approximation%
\cite{SW,pm} the equations become 
\begin{eqnarray}
\Delta _{1}(\omega )Z_{1}(\omega ) &=&\frac{\lambda _{12}\theta (\Omega -|\omega
|)}{2}\int_{-\infty }^{\infty }d\omega ^{^{\prime }}\times   \\
&&\theta (\Omega -|\omega ^{^{\prime }})|\frac{\Delta _{2}(\omega ^{^{\prime
}})}{(\Omega ^{2}+(\omega -\omega ^{^{\prime }})^{2})(\sqrt{\omega
^{^{\prime }2}+\Delta _{2}^{2}})} \nonumber \\
Z_{1}(\omega ) &=&1+\frac{1}{2\omega }\lambda _{12}\int_{-\infty }^{\infty
}d\omega ^{^{\prime }}\theta (\Omega -|\omega -\omega ^{^{\prime }}|)\times 
\nonumber \\
&&\frac{\Delta _{2}(\omega ^{^{\prime }})}{(\Omega ^{2}+(\omega -\omega
^{^{\prime }})^{2})(\sqrt{\omega ^{^{\prime }2}+\Delta _{2}^{2}})}
\end{eqnarray}%
which may be readily integrated to yield the following renormalization
behavior for $Z(\omega )$: 
\begin{eqnarray}
Z_{1}(\omega ) &=&1+\lambda _{12}\mathrm{\,\,for\,\,}\omega <\Omega  \\
&=&1+\lambda _{12}\Omega /\omega \mathrm{\,\,for\,\,}\Omega <\omega
<2\Omega  \\
&=&1+\lambda _{12}/2\mathrm{\,\,for\,\,}\omega >2\Omega 
\end{eqnarray}%
This mass renormalization behavior can then be incorporated in the previous
BCS equations yielding a natural result:%
\begin{eqnarray}
\Delta _{1}(1+\lambda _{12}) &=&\Delta _{2}\lambda _{12}\sinh ^{-1}(\omega
_{c}/\Delta _{2}) \\
\Delta _{2}(1+\lambda _{21}) &=&\Delta _{1}\lambda _{21}\sinh ^{-1}(\omega
_{c}/\Delta _{1}),
\end{eqnarray}%
reducing to Eq. \ref{BCST0} with }$\lambda _{12}\rightarrow \lambda
_{12}/(1+\lambda _{12}),\lambda _{21}\rightarrow \lambda _{21}/(1+\lambda
_{21}).$ Thus, in the linear order in $\lambda .$ {\   
\begin{equation}
\frac{\Delta _{2}}{\Delta _{1}}=\sqrt{\beta }(1+\frac{\lambda _{eff}\log
\beta }{4}+\frac{\lambda _{21}-\lambda _{12}}{2})  \label{final}
\end{equation}%
The last term is negative and always larger than the previous one
(independent of }$\beta ).$ Thus, the net effect is always opposite to what
the BCS theory predicts.{\ We have plotted up the above analytic
approximation in Figure 1 (solid line in inset) and find good
agreement for $\lambda _{eff}<0.4$, showing that the mass renormalization is
responsible for the lessening of the gap ratios with increasing coupling in
Eliashberg theory. This result might in hindsight have been expected given
that the Fermi surface with the larger gap at weak-coupling can be expected
to have larger self-energy interactions in Eliashberg theory, reducing the
gap anisotropy.  This result is also consistent with the superstrong coupling limit of equal gaps, as mentioned previously.}

Interestingly, this strong coupling effect remains operative at all
temperatures up to $T_{c},$ while the previous term in Eq. \ref{final}
vanishes at $T_{c}.$ Therefore (cf. Fig.\ref{Fig.2}) the actual gap ratio is
even closer to 1 near $T_{c}$ than at $T=0.$

Finally, {we note that the above Eliashberg results were obtained using an
Einstein spectral function for simplicity, but as indicated on the plot the
use of a typical spin-fluctuation spectrum }${[\sim \omega \Omega/(}%
\omega ^{2}+\Omega^{2}{)]}${\ does not alter the results. }

{\ Another interesting observation to be made concerns the $\Delta (0)/T_{c}$
ratios predicted by BCS and Eliashberg theory. In the conventional weak-coupling one-band
BCS theory this ratio does not depend on }$\lambda .$ This is no longer
the case in the two-band BCS with the interband coupling only. {\ In the lowest order
the reduced gaps are simply }$\Delta _{1}(0)/T_{c}=1.76\beta ^{1/4},$ $%
\Delta _{2}(0)/T_{c}=1.76\beta ^{-1/4}.$ The next order can be worked out
using Eq. \ref{BCS2}:

\begin{eqnarray}
\frac{\Delta _{1}(0)}{T_{c}} &=&1.76\beta ^{1/4}(1+\lambda \frac{%
4\log \beta -\log ^{2}\beta }{32}) \\
\frac{\Delta _{2}(0)}{T_{c}} &=&1.76\beta ^{-1/4}(1-\lambda \frac{%
4\log \beta +\log ^{2}\beta }{32})
\end{eqnarray}%
This is confirmed by numerical calculations{\ (Fig. \ref{Fig.2}): the
smaller gap ratio decreases with }$\lambda ,$ {while the other gap
increases. 
\begin{figure}[h]
{\normalsize \includegraphics[width=8cm]{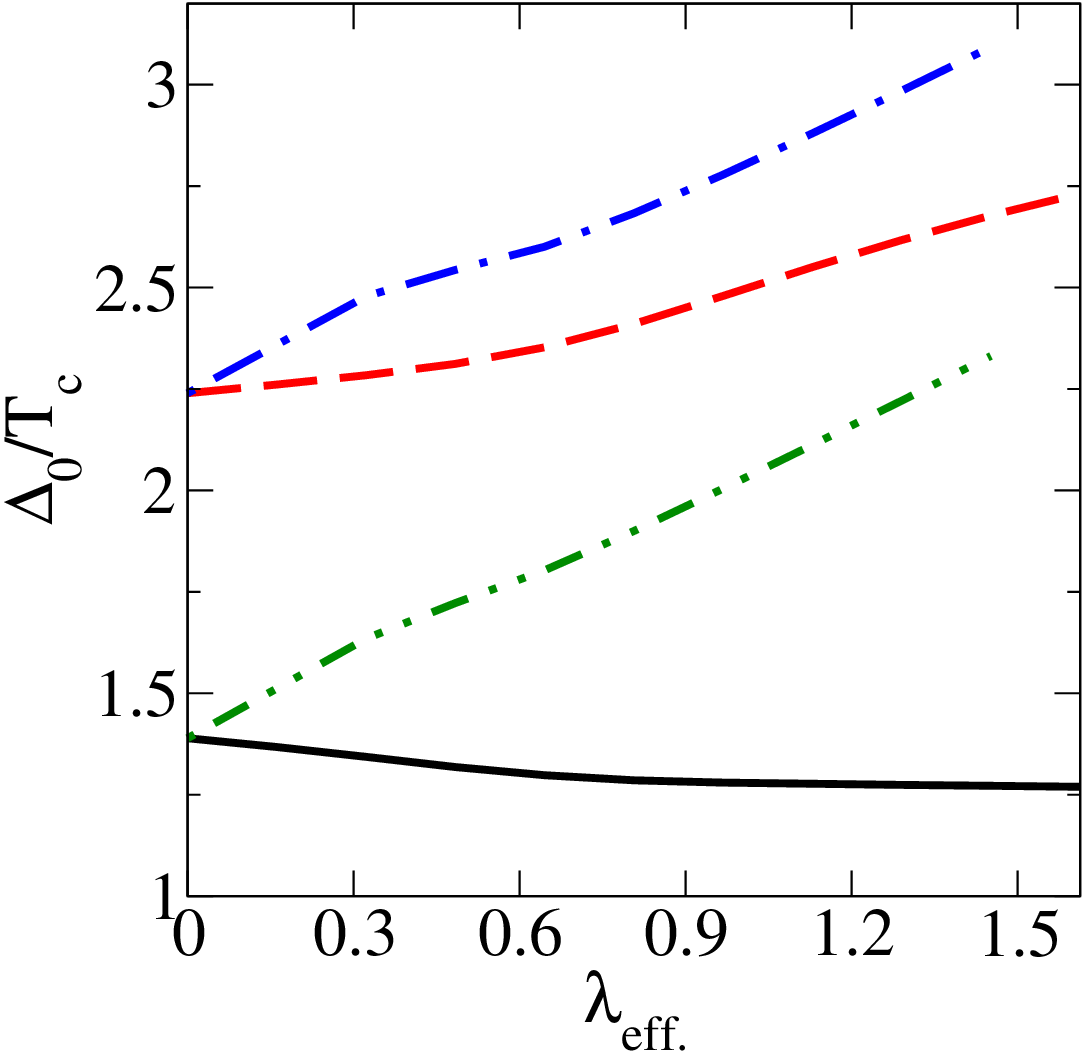} }
\caption{(color online) $\Delta(T=0)/T_{c}$ ratios are shown, as a function of the overall coupling
constant, for the BCS (solid black line and gray dashed line) and Eliashberg (dot-dashed line and double-dot-dashed line) cases.}
\label{Fig.2}
\end{figure}
Since the Eliashberg equation makes the gaps closer with increased coupling, this
odd behavior does not show up: both reduced gaps grow with }$\lambda .${\  
\begin{figure}[h]
{\normalsize \includegraphics[width=8cm]{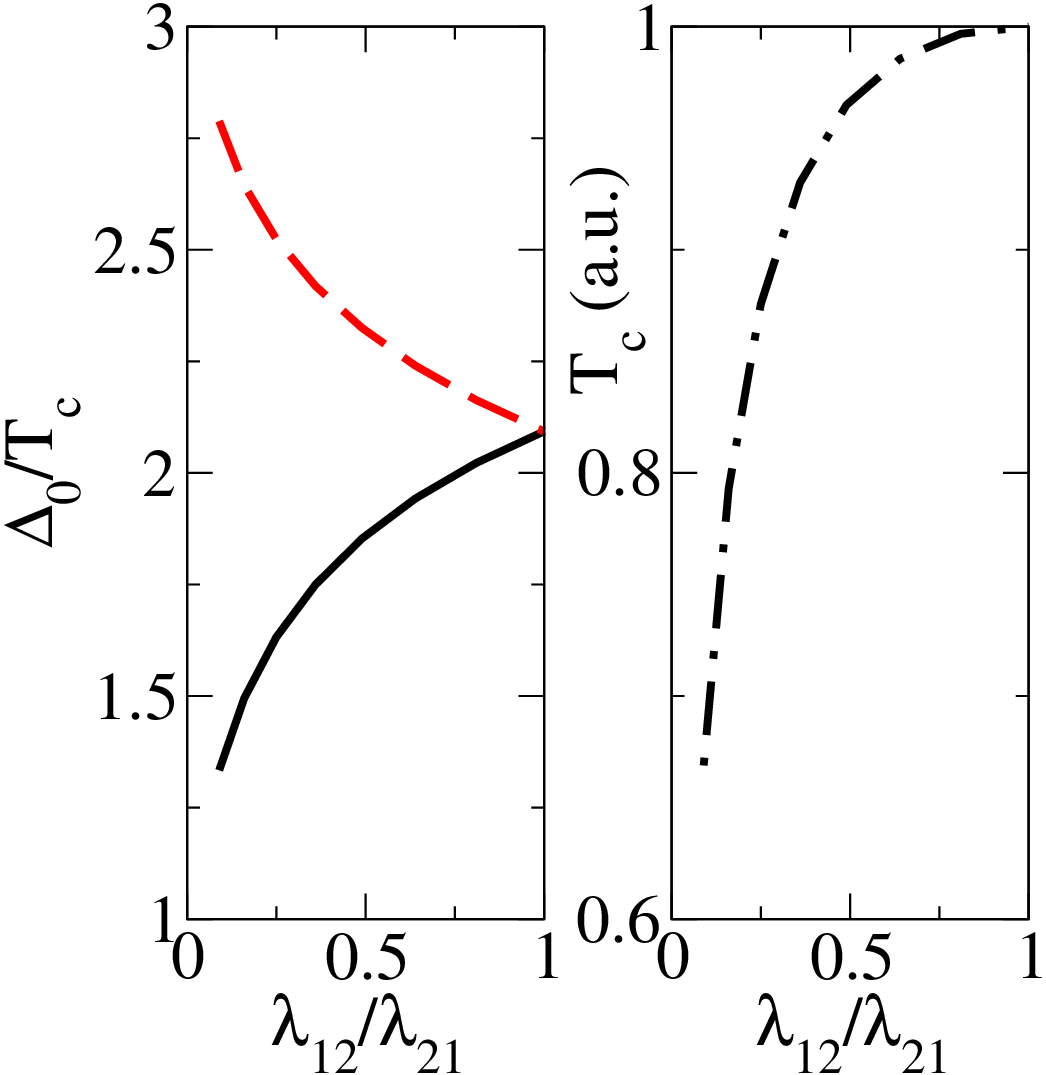} } 
\caption{(color online). (left) The behavior of the Eliashberg $\Delta(0)/T_{c}$ ratios as a function
of the ratio of coupling constants. (right) The behavior of T$_{c}$ in this
case. For both cases $\protect\lambda _{\mathrm{eff}}$ is fixed at 1.}
\end{figure}
}

{\ For completeness, we also show in Fig. 3 the behavior (in Eliashberg theory) of the reduced gaps
as a function of the DOS ratio }$N_{2}/N_{1}=${$\lambda _{12}/\lambda _{21}$%
. As might be expected, as the DOS ratio becomes very small the gap ratios move apart
appreciably. Interestingly, }${T}${$_{c}$ (shown in the right panel) is not
constant as it would be in a weak-coupling regime, but varies significantly
for coupling constant ratios far from 1. This is a result of the use of
comparatively large coupling constants on one band when the other coupling
constant is small, so that $T_{c}$ suppression due to thermal excitation of
real phonons (an effect not present in the BCS formalism) is stronger.}

{To conclude, in this work we have shown for the interband-only pairing the
two-band superconductivity is qualitatively incorrectly described by the BCS
formalism \textit{even for the weak coupling limit. }}BCS {and Eliashberg
theory predict qualitatively different behavior (as a function of coupling
constant) for such basic characteristics as the gap ratio $\alpha =\Delta
_{1}/\Delta _{2},$ as well as for the reduced gaps $\Delta /T_{c}$.  In
particular, the \textit{sign} of }$d\alpha /d\lambda $ changes from BCS to
Eliashberg theory. {\ We have found this result analytically and
numerically, by solving Eliashberg equations for model spectra. 
This finding is relevant to the superconducting pnictides where the
interband-pairing regime is believed to be realized.}

\end{document}